\documentclass[12pt]{article}
\usepackage[utf8]{inputenc}
\usepackage[T1]{fontenc}
\usepackage{amsmath,amsfonts,amssymb,amsthm}
\usepackage{lmodern}
\usepackage[shortlabels]{enumitem}
\usepackage{cite}
\usepackage{xcolor}
\usepackage[linktocpage=true, colorlinks=true, linkcolor = blue, citecolor = red]{hyperref}

\title{Universal constants \\ and natural systems of units\\in a spacetime of arbitrary dimension}

\author{A. A. Sheykin\thanks{anton.shejkin@gmail.com}, S. N. Manida\thanks{sergey@manida.com}\\
	{\it Saint Petersburg State University, Saint Petersburg, Russia}}

\date{}
\begin{document}    
	\maketitle
	\begin{abstract}We study the properties of fundamental physical constants using the threefold classification of dimensional constants proposed by J.-M. L{\'e}vy-Leblond: constants of objects (masses, etc.), constants of phenomena (coupling constants), and ``universal constants'' (such as $c$ and $\hbar$). We~show that all of the known ``natural'' systems of units contain at least one non-universal constant. We~discuss the possible consequences of such non-universality, e.g., the dependence of some of these systems on the number of spatial dimensions. In the search for a~``fully universal'' system of units, we~propose a~set of constants that consists of $c$, $\hbar$, and a~length parameter and discuss its origins and the connection to the possible kinematic groups discovered by L{\'e}vy-Leblond and Bacry. Finally, we give some comments about the interpretation of these constants. 
    \end{abstract}

\section{Introduction}
 
 The recent reform of the SI system made the definitions of its primary units (except the second) dependent on world constants, such as $c$, $\hbar$, and~$k_{B}$. This thus sharpens the question about the conceptual nature of physical constants in theoretical physics (see, e.g.,~the recent review~\cite{Hehl2019}). This question can be traced back to Maxwell and Gauss, and~many prominent scientists made their contribution to the discussion of it. The~analysis of fundamental constants could help us understand the structure of the underlying physical theories, as~well as explain this structure. The~Gamow--Ivanenko--Landau--Bronstein--Zelmanov cube of physical theories (or simply $Gc\hbar$ cube) is possibly the most famous example of such analysis. It has been a~popular educational and methodological tool for many decades~\cite{barrow2003constants,Padmanabhan2015} in various forms and generalizations~\cite{Faddeev1991,Okun2002,Lajzerowicz_2018}, even though some of the vertices of such a~cube are still~empty.
 
 The physical constants can also serve more practical purposes. Since the end of the XIX Century, physicists have been trying to employ various sets of them to invent systems of units that could be ``natural'' for one or another theory~\cite{tomilin2000}. When constructed, such systems of units allow one to make some qualitative claims about the characteristic behavior of the corresponding physical theory, e.g.,~``the characteristic length of quantum gravity is $10^{-33}$ meters, so measurements are impossible beyond this scale'' \cite{Gorelik_2005} or ``in classical electrodynamics, the~characteristic radius of an electron is $10^{-15}$ meters, so at this scale, renormalization might be needed'' \cite{baez_lengths}.

 The possibility to build up characteristic scales out of physical constants and to compare them to each other leads to Dirac's large numbers hypothesis and its numerous variations~\cite{ray2019,barrow2003constants}. Another~related issue, which arises when one considers some set of physical constants, is the fact that their numerical values seem to be not so random as one might expect. The~most common implication of this fact is the so-called anthropic principle, which postulates that the values of physical constants in the observable world must provide the existence of a~certain observer (see, e.g.,~the recent review~\cite{Naumann2017} and the references therein). The~discussion of this principle has revived in recent years in connection with the concept of a~multiverse, which is a~feature of some inflationary theories~\cite{Heller2019,Dabrowski2019}.

 However, in~the construction of such systems, it is often overlooked that the constants that serve as their base might have different degrees of ``fundamentality''. Furthermore, there are many ways to define what constant is more fundamental than another, as~well as many opinions on that~\cite{Duff_2002}.
 
 Several classification schemes were developed, in~which the constants of nature are divided into groups according to their appearance in physical laws. One of such schemes was proposed by T. Afanassjeva-Ehrenfest~\cite{aehrenfest1926} as early as in 1926. Another one was discussed by Einstein and his student Ilse Rosenthal-Scheider~\cite{Ilse,Ilse2} (see also~\cite{barrow2003constants}). In~these classifications, all types of constants were considered, both dimensional and dimensionless. In~the present paper, however, we want to concentrate our attention on the classification of dimensional physical constants presented by \mbox{J.-M. L{\'e}vy-Leblond}~\cite{LL1977}. This is based on their role in the laws and theories (a similar classification was also discussed in~\cite{stepanov1999fundamental}). According to L{\'e}vy-Leblond, three possible types of constants~are:
 \begin{enumerate}[leftmargin=2.1em,labelsep=4mm,label={\Alph*}.]
 \item the characteristics of the objects, such as the masses of elementary particles,
 \item the characteristics of the phenomena (i.e.,~interactions), such as elementary
 charge or the gravitational constant,
 \item universal constants, such as $c$, $\hbar$, and~$k_{B}$. 
 \end{enumerate}
 
 Most of the natural systems of units contain quantities of Type A (such as the electron mass in the electrodynamic system of units), or~Type B (such as gravitational constant $G$ in Planck units), or~both. One might ask about the possible consequences of this inhomogeneity and the properties of such systems of~units. 
 
 It turns out that dimensional analysis can help draw some distinction between constants of different types, as~well as shed some light on their~properties.
 
 Another interesting question is whether it is possible to construct a~set of constants that would be universal in the above sense, i.e.,~contain only those constants that can be considered as universal~ones.
 
 This article, which is intended to serve methodical and pedagogical purposes, is organized as follows. In~Section~\ref{Sec2}, we examine the properties of constants of interaction, namely $G$, $e$, and~Yang--Mills coupling constants $g_s$, $g_w$, and~derive their dimensionality in the spacetime with $n$ spatial directions. In~Section~\ref{Sec3}, we generalize some well-known natural systems of units, such as Planckian and field-theoretic ones, to~the case of $n$ dimensions. It turns out that the dimensionality of space affects the explicit form of their base units. In~Section~\ref{Sec4}, we construct a~set of ``fully universal'' constants, the~dimensions of which are independent of the dimension of space. Besides~$c$ and $\hbar$, this set contains some fundamental length. We discuss the origin of these constants and their connection with the structure of the most general kinematic group~\cite{bacry}. In~Section~\ref{Sec5}, we review the history of the notion of fundamental length and give some concluding remarks about its~interpretation.
 
 \section{Constants of the Fundamental~Interactions}\label{Sec2}
 
 Let us consider the expression for the Newtonian gravitational force between two
 point masses in $n$ spatial dimensions:
\begin{align}
 F = G_n\frac{m_1 m_2}{r^{n-1}}, \label{newton}
 \end{align} 
 which, as~is well known, follows from the Gauss theorem and the assumption that
 gravitational force is long range~\cite{ehrenfest_3D}. If~none of the other
 fundamental constants are present, the~dimensionality of $G_n$:
\begin{align}
 [G_n]=M^{-1} L^{n} T^{-2}, \label{G}
 \end{align} 
 itself gives us little information about the physics in an $n$-dimensional
 space. Things become interesting when $c$ enters the game
 (for a~discussion of the gravitational force in $n$ dimensions in the presence of other constants, see, e.g.,~\cite{Barrow_2014,barrow2020maximum}). For~instance, a~simple calculation shows that, if~one considers a~point body of mass $M$ in a~universe with two spatial dimensions, it turns out that it is impossible to construct the quantity with the dimension of length out of $M$, $c$, and~$G_2$ (i.e.,~2D gravitational radius). Therefore,~the~metric of spacetime $g_{\mu\nu}$, which ought to be a~dimensionless function of the spacetime coordinates $x$, in~fact, cannot depend on them, because~there is no dimensionless combination containing $x$. The~force between two particles thus is equal to zero, and~it seems that the Formula \eqref{newton}, which was the starting point of this consideration, is no longer valid in 2~+~1 dimensions and should not be used for the derivation of the dimensionality of $G_n$. However, in~the relativistic framework, the~dimension of $G_n$ in the $n+1$-dimensional gravity can also be deduced from the gravitational part of Einstein--Hilbert
 action:
\begin{align}\label{EH} 
S=- \frac{1}{2\kappa_n}\int d^{n}x dt R,
\end{align}
where $\kappa_n=8\pi G_n/c^4$ is the Einstein gravitational constant and $R$
is the scalar curvature of the spacetime. Since the dimension of $S$ is always
$[S]=ML^2T^{-1}$ and $[R]=L^{-2}$, Equation \eqref{G} with arbitrary $n$ follows
from it immediately. As~a~side result, we obtain that for classical general relativity with
point masses, the~dimension 2~+~1 is critical: the metric of spacetime with point
masses is flat---a~well-known fact~\cite{Staruszkiewicz,deser_3D}, but~obtained in a~very straightforward~way. 
 
 The dimension of elementary charge $e$, which can serve as the electromagnetic coupling constant, can be derived from the Coulomb law:
\begin{align}
 F=\frac{e^2}{r^{n-1}} \Rightarrow [e]=\sqrt{ML^n} T^{-1}.
 \end{align}

 Since the two remaining interactions, namely the strong and weak ones, are short range, the~Gauss theorem cannot be directly applied to obtain the dimension of their coupling constants. We know, however, that these interactions are described by non-Abelian gauge fields, also known as Yang--Mills fields. Let us write the action of the Yang--Mills field interacting with some charged point particles in an arbitrary dimension:
\begin{align}\label{YM} 
 S= \int d^{n}x dt \ \text{Tr} \ F_{\mu\nu} F^{\mu\nu} + \text{Tr} \ \frac{g}{c} \int A_\mu dx^\mu,
 \end{align}
 where $g$ is a~coupling constant, $F_{\mu\nu} = \partial_\mu A^a_\nu -\partial_\nu A^a_\mu + g [A_\mu, A_\nu] $, $\mu,\nu = 0\ldots n$, and~each of the components of $A_\mu$ is a~matrix whose size depends on the gauge group that is considered. 
 The~point particles with Yang--Mills charges may seem peculiar, but~as a~toy model, they serve well~\cite{ym2011}; without~them, some lengthy talk about covariant derivatives would be unavoidable.
 Comparing~the dimensions of both terms to the standard dimension of action, we get:
\begin{align}
 M L^2 T^{-1} = L^n T [A]^2 L^{-2} = [g] L^{-1} T [A] L \ \Rightarrow \ [g] = \sqrt{ML^n} T^{-1}.
 \end{align}
 
 The dimension of coupling constant $g$ turns out to be the same as $e$, which is not surprising because the EM field is a~particular case of the Yang--Mills~field.
 
 As a~result, we obtain the dimensions of all four coupling constants in arbitrary dimensions: $[G_n]=M^{-1} L^n T^{-2}$ and $[e]=[g_s]=[g_w] = \sqrt{ML^n} T^{-1}$. As~can be seen, all of them depend on $n$. It~is tempting to speculate that this occurs due to the local character of all four interactions. Indeed,~the~actions of local theories can be represented as integrals of Lagrangian density. Since the integration measure depends on $n$, while the dimension of the action does not, one can conclude that Lagrangian density itself (and therefore, coupling constants) must depend on $n$ to compensate the dependence of the measure. Note that this is the main formal difference between constants of Types A and B: as we saw, the~dimension of Type B constants depends on the characteristics of the space, whereas the dimension of Type A constants, in~general, does~not.
 
 \section{Natural Units and the Dimension of~Space}\label{Sec3}
 We are now ready to construct the $n$-dimensional generalization of natural units. Let us start with Planck units, which consist of $G_n, \hbar$, and~$c$:
\begin{align}
 l_{P}=\left(\frac{\hbar G_n}{c^3}\right)^{\xi}, \quad m_{P}
 = \frac{\hbar}{c\,l_{P} }, \quad t_{P}=\frac{l_{P}}{c}, \quad \xi=\frac{1}{n-1}.
 \end{align}
 
 In this case, the~dimension $n=1$ is critical, which is possibly connected to the fact that in $1+1$-dimensional spacetime, the~EH gravitational action \eqref{EH} is trivial (more precisely, it equals the so-called Euler characteristics of the spacetime, which is a~constant number) \cite{Boozer_2008,fletcher2018,stojkovic2014}.
 
 As we stated in the Introduction, we want to stress that the expressions for Planckian units depend on $n$ since $G_n$ does. It is not surprising if the constant $G$ is regarded as belonging to Type B, i.e.,~as the coupling constant and not as a~fundamental characteristic of a~physical theory. However, it should be noted that the type of a~certain constant might change over time. It is possible, though, that after the completion of quantum gravity, $G$ would be promoted to Type C (as, e.g.,~the electron charge $e$ was promoted from Type A to Type B after the completion of QED). For~now, however, $G$ must be associated with Type B.
 Therefore, the~Planck scale physics is different in different dimensions, and~we cannot make any Planck-based claim about the nature of spacetime without specifying its~dimension. 
 
 This also occurs in the case of other constants of interaction. For~instance, we could attempt to construct some natural units using $c$, $\hbar$, and~the coupling constant of one of the other interactions (we~denote it as $g_n$). Such a~system corresponds to a~quantum theory of some field, so its units can be called QFT units.
 We obtain that:
\begin{align}
 l_{QFT}=\left(\frac{g_n^2}{\hbar \,c}\right)^{\zeta}, \quad m_{QFT}
 = \dfrac{\hbar}{c\,l_{QFT} }, \quad t_{QFT}=\dfrac{l_{QFT}}{c}, \quad \zeta = \frac{1}{n-3}.
 \end{align}
 
 Again, the~dependence on $n$ is present. Note that our space with three spatial dimensions represents the critical case here: at $n=3$, there is no characteristic length (and therefore energy) scale corresponding to the interaction, so it becomes renormalizable (when $n<3$, it is super-renormalizable, and~when $n>3$, it is non-renormalizable) \cite{Delamotte_2004,Olness_2011}. There are also two remaining possibilities to construct a~system of units out of constants of Types B and C. The~first one is Stoney units~\cite{tomilin2000} based on $e, G$, and~$c$, whose critical dimension is $n_c=2$, and~the second one is some unnamed peculiar system based on $G, e$, and~$\hbar$ with $n_c=4$. It is also worth noting that $G$, $e$, $c$, and~$\hbar$ allow one to construct a~dimensionless constant that is independent of $n$: the Barrow--Tipler invariant~\cite{barrow1986anthropic,Barrow_2014}:
\begin{align}
G^{(3-n)/2} e^{n-1} c^{n-4} \hbar^{2-n}=\text{const}. 
 \end{align}

 It is possible to construct many other systems of units, such as electrodynamic ($m_e$, $c$, and~$e$) and atomic ($m_e$, $\hbar$, and~$e$) ones, which can also be generalized to the case of arbitrary dimension. While~this procedure can shed some light on a~theory that possesses such constants, it also shows that none of the theories, in~which constants of Type B are present, allow the construction of natural units that are independent of the number of spatial dimensions. The~usage of Type A constants, such as the masses of particles, allows one to construct $n$-independent systems (e.g., $m_p, c, \hbar$) and $n$-independent dimensionless constants (such as $Gm_e m_p/e^2$) \cite{Barrow_2017}. However, one then has to deal with the arbitrariness of the choice of mass scale: why one mass and not the other? One possible reasonable choice might be the Higgs mass~\cite{Wilczek2007}, since it is the Higgs field that gives the mass to all other particles. However, if~it turns out that there is more than one Higgs boson, then such a~choice would be~inconvenient.
 
 Therefore, a~question arises: Are there enough constants of Type C to construct a~system of units out of them?
 
 The most popular answer is no, since in the present state of theoretical physics, only $c$ and $\hbar$ are recognized as Type C constants, and~the number of base units in mechanics is three. Another~example of a~physical constant that could belong to this type is Boltzmann constant $k_B$~\cite{LL1977}. However,~to~introduce it, one has to define the base unit of temperature, raising the number of base units. Therefore,~one~constant is still missing. If~it is not a~mass of any particle, we could expect that it might be some universal length (or, equivalently, a~time interval). It turns out that there is a~natural way to include it in the theory alongside $c$ and $\hbar$. We discuss it in the next~section.
 
 \section{Kinematic Groups and Curvature of~Space}\label{Sec4}
 In this section, we are going to show that one candidate for the role of the third constant is the curvature radius of the~spacetime. 
 
 In classical mechanics, three quantities with a~nontrivial dimension can be adopted as base units. They are usually chosen to have units of length, mass, and~time, although~other choices are possible (and even were recently advocated) \cite{Hehl2019}. Let us consider the Galilean group, which is a~symmetry group of classical mechanics. A pedagogical explanation of the properties of the Galilean group and its generators can be found, e.g.,~in~\cite{0810.4637}. The~relation of the Galilean group and other groups to the structure of dimensional quantities was discussed in~\cite{Carinena1981,Carinena1985}. Its algebra of generators in $n+1$ dimensions has the form (we omit indices and Kronecker deltas for the sake of simplicity) \cite{Gomis1978}:
\begin{align}\begin{split}
 &[M,P]=P, \ \ \ \ [M, K]= K, \ \ \ [M,M]= M,\\
 &[H,P]=0, \ \ \ \ \ [H,K]= -P, \ \ [H,M]=0, \\
 &[P,P]=0, \ \ \ \ \ \, [K,K]=0, \ \ \ \ \  [P,K]=0,\end{split}
 \end{align}
 where $M$, $P$, $K$, and~$H$ are the generators of rotation, spatial translation, boost, and~time translation,~respectively.
 
 It is known that this can be deformed in many ways to obtain various generalizations of kinematics~\cite{bacry}. It can be shown~\cite{Herranz_2007} that after two steps of such deformations, the~(anti)-de Sitter group appears. The~commutation relations of its generators are:
\begin{align}\label{ads}\begin{split}
 &[M,P]=P, \qquad \ [M, K]= K, \qquad \ [M,M]= M,\\
 &[H,P]=\pm\dfrac{c^2}{R^2} K, \ [H,K]=-P, \ \ \ \ \ [H,M]=0, \\
 &[P,P]=\mp \dfrac{1}{R^2} M, \ [K,K]=-\dfrac{1}{c^2}M, \ \, [P,K]=- \dfrac{1}{c^2} H,
 \end{split} 
 \end{align}
 where $R$ is a~parameter with the dimension of length, and~the upper sign in the commutators corresponds to the AdS case and the lower to the dS one. As~was shown in~\cite{Ballesteros_2007}, both the Galilean and Poincare algebra of $n+1$-dimensional flat spacetime can be considered as contractions of an algebra of a~certain space with constant curvature. In~the case of Poincare to Galilean group contraction, such~curvature is $-1/c^2$ and can be considered as a~curvature of the space of velocities. In~the case of the (A)-dS to Poincare group, the~curvature of spacetime is $\mp 1/R^2$. 
 
 As can be seen, alongside $c$, we have another deformation constant $R$. Moreover, in~the case of anti-de Sitter space, there is a~so-called $R$-$c$ duality~\cite{Manida_2011}, so the contraction w.r.t. either of the two parameters of the anti-de Sitter group can lead to the Poincare group, and~the roles of these parameters are completely analogous. 
 We can see this if we replace $H$ with $c^2\hat{H}$ in some of the commutation relations (\ref{ads}):
\begin{align}\label{ads2}\begin{split}
 &[\hat{H},P]=+\dfrac{1}{R^2} K, \ [\hat{H}, K]= - \dfrac{1}{c^2} P, \ \ [\hat{H},M]=0, \\
 &[P,P]=- \dfrac{1}{R^2} M, \ [K,K]=-\dfrac{1}{c^2}M, \ \, [P,K]=- \hat{H}.
 \end{split} 
 \end{align}
 
 We get the standard Poincare algebra in the limit $R\to \infty$ in (\ref{ads2}), and~we get the ``second'' Poincare algebra in the limit $c\to \infty$, where the roles of the boost and the spatial translation are swapped and the sign of the time translation generator~changed.
 
 It is also worth noting that, as~was discovered by V. A. Fock~\cite{Fock-kniga}, the~most general form of the transformation between the coordinates in the two inertial frames is the fractional linear transformation, which contains both $c$ and $R$ as parameters~\cite{Manida1999}. The~physics in the theory with fractional linear transformations was also investigated in~\cite{Stepanov_2000}.
 
 Finally, if~one considers the algebra of quantum operators corresponding to \eqref{ads}, the~Planck constant $\hbar$ appears on the right side of the commutators of operators. It makes its appearance because the dimension of operators is governed by canonical commutation relations and cannot be arbitrary. The~role of $\hbar$ as a~deformation constant was discussed, e.g.,~in~\cite{Faddeev1991}. 
 
 We can conclude that three constants with nontrivial dimensions have the same ``deformation'' origin, and~their properties are independent of the number of space dimensions, so one can construct a~full set of units on the basis of them. Such units are closely related to the de Sitter units that were discussed in~\cite{Barrow_2014}; see also~\cite{Dadhich_2017}. The~physical sense of these three constants is also similar since all of them allow one to connect different physical notions (a remarkable feature of Type C constants) \cite{LL1977}. Indeed, $c$ connects the notions of space and time, $\hbar$ the coordinate and momentum (or energy and time, etc.), and~$R$ the length and angle. Therefore, all these constants are, in~some sense, constants~of~relativity.

 There is no need to discuss the properties of $c$ and $\hbar$ here. The~properties of universal length $R$, however, deserve some study. We discuss the notion of fundamental length\footnote{In the context of this paper, the~terms ``fundamental length'' and ``universal length'' are, strictly speaking, not synonymous. Indeed, the~Planck length is, without~any doubt, fundamental, as~it is constructed of fundamental constants. However, it is not universal, since one of these fundamental constants, $G$, is not universal in the sense of the above-mentioned classification. In~many cases, however, such a~distinction is difficult to draw, so we will nevertheless use both terms as synonyms.} and its interpretations in the last~section. 
 
 \section{Discussion}\label{Sec5}

 In this article, we stated that Planck units, which are widely assumed to be related to the properties of spacetime, are not invariant with respect to the change of the dimension of space. This occurs due to the inhomogeneity of the set of physical constants, on~which these units are based: $G$ is not a~universal constant in the sense of the L\'evy-Leblond classification. Replacing $G$ with the universal length $R$, we obtain a~homogeneous set and thus have such invariance. The~methodological reason for the assumption of universality of $R$ lies in the properties of a~possible kinematic group of spacetime, which might contain a~constant length parameter $R$ alongside a~constant velocity parameter $c$, both of which can be considered as curvatures of a~certain~space. 
 
 The concept of fundamental length has a~very long story, and~there are two directions of investigation: a~small fundamental length and a~large one. Since the existence of a~small fundamental length could alter the physics on small scales (and high energies), the~inclusion of minimal length was initially discussed in the framework of quantum physics~\cite{1001.0050}. Later, these studies gave rise to more specific theories that deal with the concrete realization of fundamental length. The~examples of those are non-commutative geometry (its application to the problem discussed here can be found, e.g.,~in~\cite{1111.5576} and the references therein) and double-special relativity (the connection of double-special relativity
 with the deformation of operator algebra was discussed, e.g.,~in~\cite{Calisto2005}). 
 
 On the other side, the~existence of fundamental length is one of the consequences of the Kaluza--Klein theory (its basic overview can be found in~\cite{Chyba}; for a~more detailed account, see, e.g.,~\cite{gr-qc/9805018}). If~the description of electromagnetism (or quantum mechanics~\cite{rumer1959}) in the Kaluza--Klein framework is desired, this length ought to be very small. The~KK theory can also be treated as one of the predecessors of string theory~\cite{rickles}, in~which some small fundamental scale $\alpha'$ is also present. In~the development of string theory, in~turn, various brane theories appeared. In~many of them, the~fundamental length is assumed to be large (as in the Randall--Sundrum model, where it is related to the curvature of five-dimensional bulk spacetime, or~in the Arkani-Hamed--Dimopoulos--Dvali 
 model, where it plays the role of the compactification radius) \cite{branes_ajp}.

 The existence of a~large fundamental length, on~the contrary, was initially discussed in the context of general relativity and cosmology. For~instance, the~two most popular early cosmological models, namely Einstein and de Sitter ones, both have certain characteristic length scales (this is the reason why Friedmann called them ``cylindrical'' and ``spherical'' universes, respectively) \cite{Friedman1999}. In~the framework of general relativity, the~cosmological constant introduced by Einstein could be treated either as a~spacetime curvature or as a~``vacuum energy'' \cite{Straumann2002}. Therefore, the~question of the existence of a~large fundamental length had soon become a~part of the so-called cosmological constant problem (although~there were attempts to connect the cosmological constant with some ``atomic'' length; see~\cite{Anderson1971}). A~brief exposition of the quantum side of this problem can be found in~\cite{adler1995}. For~a~historical review of the cosmological side of the problem, see~\cite{ORaifeartaigh2017} and the references therein. Since in our universe, the~quantity $R$, which is discussed in this paper, has to be, by~construction, quite~large (in fact, so large that we cannot or barely can notice its presence), we can conclude that its role is similar to the role of the cosmological~constant. 
 
 We want to stress here that we are not claiming that the quantity $R$ must have the value corresponding to the observed density of dark energy. Some researchers still make attempts to solve the cosmological constant problem using similar kinematic considerations (see, e.g.,~\cite{Lev_2020}). However,~such~attempts, in~our opinion, are hardly convincing. This problem is not merely about the value of spacetime curvature, but~also about its relation to microphysics. Moreover, the~nature of the dark energy (which could be treated as an effect of the presence of the cosmological constant) remains insufficiently clear, especially due to the data that appeared in the last few years~\cite{valentino2020cosmic,DiValentino2020}. The~reduction of dark energy to the cosmological constant leads to another problem, namely the coincidence problem: why is the value of vacuum energy (i.e., the~cosmological constant) so close to the value of the energy of other matter, which is supposed to be independent of it~\cite{Velten2014}? Finally,~the~fundamental length $R$ is somehow different from fundamental velocity $c$ and fundamental action $\hbar$. We cannot ask why $c$ is so big or $\hbar$ is so small (with the assumption that they are true constants and are not affected by any dynamical process), as~we have nothing nearly as fundamental with which to compare them. However,~we~can ask why $R$ (if it exists) is so big in comparison to the scales of all fundamental interactions, as~well as why the constant spacetime curvature produced by $R$ is so small in comparison with the non-constant parts of spacetime curvature produced by a~matter of a~different kind. Such~a~question, as~was mentioned in the Introduction, would lead to the famous large numbers hypothesis and its variations (\cite{ray2019}; see also ch. 23 of~\cite{Harrison}) and can not be solved without some assumptions on dynamics, while in this paper, we discuss only the kinematic properties of spacetime. For~the same reason, we also omit the discussion of other constants of a~mechanical origin, namely the maximal force (an example of such an approach can be found in~\cite{Barrow_2014,Barrow_2020}) and acceleration~\cite{Torrome_2018}.
 
 In other words, we do not know if $R$ exists, but~if it {does} exist, it could form, together with $c$ and $\hbar$, some set of universal constants (or Type C constants, or~constants of relativity), and~the corresponding system of units would be independent of the number of spatial dimensions. The~search for such a~system is the first main goal of this~paper.
 
 Secondly, we want to note the inhomogeneity of all other known systems of natural units, especially Planckian ones, and~conclude that due to this, they are not so suitable for methodological considerations of the structure of physical theories as is widely assumed. The~constants of interactions $G$, $e$, $g_s$, and~$g_w$ and ``constants of relativity'' $c$, $\hbar$, and~$R$ (and $k_{B}$, whose role in this context was discussed in~\cite{LL1977}) play drastically different roles in physical theories, and~this circumstance needs to be especially underlined in the discussion of various theories and corresponding systems of~units.
 
    {\bf{Acknowledgments}}. The authors are grateful to I.~Ado, E.~Anikin, D.~Grad, D.~Kalinov, M.~Markov, O.~Novikov, S.~Paston, and K.~Pavlenko for useful discussions. The authors also want to thank J.~Barrow for pointing out many useful references. The work of A. S. was supported by RFBR grant 20-01-00081.

\end{document}